\documentclass[a4paper,11pt]{article}

\usepackage{jinstpub} 

\usepackage{wrapfig}
\usepackage{subcaption}
\usepackage{cleveref}

\title{Incorporating Inelasticity Reconstruction into Neutrino Mass Ordering Studies with IceCube }

\collaboration[c]{on behalf of the IceCube Collaboration$^*$\note[*]{Full author list and acknowledgments are available at \href{https://icecube.wisc.edu/collaboration/authors/\#collab=IceCube&date=2025-08-25&formatting=web}{icecube.wisc.edu}.}}

\author{J. H. Peterson}
\affiliation{University of Wisconsin - Madison,\\Madison, WI 53706, USA}
\author{and M. Jacquart}
\affiliation{University of Copenhagen,\\Copenhagen, Denmark}

\emailAdd{josh.peterson@icecube.wisc.edu}
\emailAdd{marc.jacquart@icecube.wisc.edu}

\abstract{Earth's matter affects the oscillation of atmospheric neutrinos and antineutrinos differently depending on the neutrino mass ordering (NMO). As more neutrinos than antineutrinos are expected to be detected in the IceCube detector, this matter effect can be used to probe the NMO. The fraction of energy transferred to the nucleon during a neutrino interaction, known as the inelasticity, has a different distribution for neutrinos and antineutrinos because of their opposite chirality. This can in theory be used to statistically separate neutrinos from antineutrinos, but hasn't been exploited in IceCube DeepCore analyses yet. To this end, two new inelasticity reconstructions were developed using a graph neural network and an ensemble of two-dimensional convolutional neural networks. This presentation discusses the development and performances of these reconstruction algorithms. The inelasticity is then used as a fourth observable, along with the particle energy, direction and flavor, to calculate new NMO sensitivities and determine the impact of adding the inelasticity in the measurement of the NMO with the IceCube DeepCore and upcoming IceCube Upgrade detectors.}

\proceeding{Presented at the 32nd International Symposium on Lepton Photon Interactions at High Energies\\
  Madison, Wisconsin, USA\\
  25-29 August, 2025}

\notoc
\begin{document}
\maketitle
\flushbottom

\section{Introduction}
\label{sec:introduction}

The IceCube Neutrino Observatory consists of 86 support and readout cables, called "strings", each holding 60 digital optical modules (DOMs) that detect Cherenkov radiation from neutrino interactions in the Antarctic ice \cite{IceCube_Detector}.  IceCube DeepCore (IC86) and the upcoming additional 7 strings of the IceCube Upgrade (IC93) are more densely instrumented infills to lower the detection energy threshold from hundreds of GeV to about $5\thinspace$GeV and $3\thinspace$GeV, respectively \cite{FLERCNN, QUESO}.

Atmospheric muon neutrinos passing through the Earth can oscillate into tau neutrinos before detection.  Neutrinos passing through Earth experience a potential from atomic electrons that alters their oscillation.  The resonance of this matter effect will only affect neutrinos or antineutrinos depending on whether the neutrino mass ordering (NMO) is normal or inverted, respectively.  Since the IceCube Neutrino Observatory expects to detect more neutrinos than antineutrinos we can look at the intensity of these matter effect signals to determine the NMO \cite{FLERCNN_NMO}. 

Inelasticity, $y$, is the fraction of neutrino energy that is deposited into the hadronic cascade during its scattering with the nucleon.  Because of their chirality and the weak interaction structure, scattering of antineutrinos with a high inelasticity are suppressed compared to neutrinos \cite{Devi_2014}.  Measuring the inelasticity would allow a partial statistical separation between neutrinos and antineutrinos, thus increasing the sensitivity to the NMO.  These proceedings summarize the latest work in exploring this possibility.

\section{Inelasticity Reconstruction}
\label{sec:y_reco}

New Machine-learning-based inelasticity reconstruction algorithms were developed for both IC86 and IC93 detector configurations.  Both algorithms were trained on Monte Carlo simulations of muon neutrino charged current interactions, called track events.  This is because the visible outgoing muon can be spatially separated from the hadronic cascade, which allows us to reconstruct the inelasticity.  The algorithms were then tested on other neutrino flavor events (cascades), where the lack of muon track should give an inelasticity reconstruction close to one.

\begin{table}[h]
\centering
\caption{Differences in how the convolutional neural networks were trained for IC86 inelasticity reconstruction.}
\begin{tabular}{ |c|c|c|c|c|c| } 
 \hline
  CNN & Selection Level & Energy Spectrum & Lower Energy Cut & Containment Radius & Loss \\
 \hline \hline
  1 & level 3 & nominal & 3 GeV & 100 m & Beta \\ 
 \hline
  2 & level 6 & flat & 5 GeV & 200 m & L1 \\ 
 \hline
  3 & level 6 & flat & 30 GeV & 200 m & L1 \\ 
 \hline
\end{tabular}
\label{cnn_diffs}
\end{table}

For IceCube DeepCore, three convolutional neural networks were trained to measure inelasticity utilizing the architecture described in reference \cite{CNN}.  Table \ref{cnn_diffs} summarizes the differences in how the networks were trained.  Flat energy spectra and higher energy cuts produced better results for high-energy events.  Nominal energy spectra and lower energy cuts produced better results for low-energy events.  Using a tighter cut on the neutrino interaction vertex in DeepCore, having the network output a beta probability distribution instead of a single inelasticity, and using lower levels of the event selection described in \cite{FLERCNN} for larger training samples all improved the reconstruction performance \cite{FLERCNN}.

The output of the three networks are then fed into a boosted decision tree (BDT) to obtain the final reconstructed inelasticity.  This BDT is trained with a flat inelasticity distribution using the final level event selection described in \cite{FLERCNN}.  This was found to work better than using a single CNN or using other aggregating methods like averaging.

For IC93 the graph neural network DynEdge model \cite{GNN} is used with the DOMs positions and detected photons time of arrival as input.  It was trained using the IceCube Upgrade final level simulation events sample with a flat inelasticity distribution and with a log-cosh loss function \cite{QUESO}.

\begin{figure}[htbp]
\centering
\includegraphics[width=.45\textwidth]{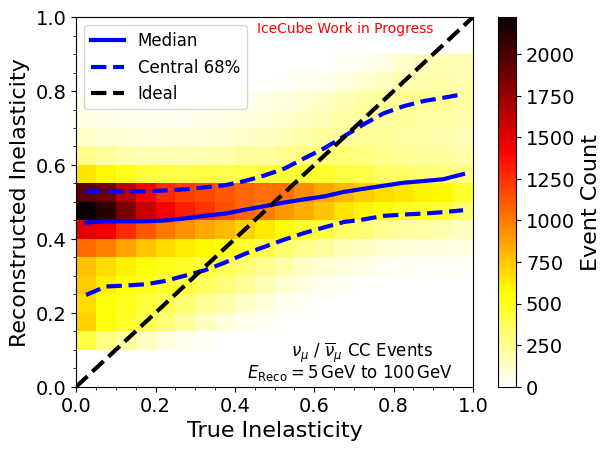}
\includegraphics[width=.45\textwidth]{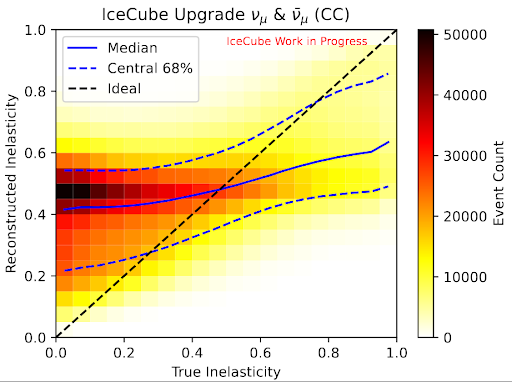}
\caption{\label{fig:y_recos} True versus reconstructed inelasticity for IC86 (left) and IC93 (right).}
\end{figure}

Figure~\ref{fig:y_recos} shows a comparison of the true inelasticity of track events to the reconstructed inelasticities.  The notable flat distributions near $y_{\text{reco}}=0.5$ are comprised of events below 20 GeV that have been reconstructed near the mean inelasticity of the training set.  As the true energy increases the inelasticity reconstructions improve in quality.  The improvement seen in the IC93 plot compared to the IC86 one is mainly due to the reconstruction improvement of lower energy events.  Both reconstructions were found to have limited performance in the energy range relevant for matter effects in oscillations, but there was enough performance to make it worth implementing them in the NMO analysis.

\begin{table}[h]
\centering
\caption{Binning used for IC86 (top) and IC93 (botton) NMO sensitivity calculation.}
\begin{tabular}{ |c|c| } 
 \hline
  Observable & Binning (IC86) \\
 \hline
  PID and $y_{\text{reco}}$ & y<0.47 and PID [0.0, 0.55, 1.0], y>0.47 and PID [0.0, 0.25, 1.0] \\ 
 \hline
  $E_{\text{reco}}$ &  5GeV to 100GeV, 12 log$_ {10}$ bins\\ 
 \hline
  $\cos(\theta_{\text{zenith}})_{\text{reco}}$ & -1.0 to 0.0, 10 linear bins \\ 
 \hline \hline
   Observable & Binning (IC93) \\
 \hline
  PID and $y_{\text{reco}}$ & PID < 0.3, PID [0.3, 0.85, 1.0] $\times$ y [0.0, 0.46, 1.0] \\ 
 \hline
  $E_{\text{reco}}$ &  3GeV to 300GeV, 12 log$_ {10}$ bins\\ 
 \hline
  $\cos(\theta_{\text{zenith}})_{\text{reco}}$ & -1.0 to 0.0, 10 linear bins \\ 
 \hline
\end{tabular}
\label{binning}
\end{table}

\section{Neutrino Mass Ordering with Inelasticity}
\label{sec:nmo_w_y}

We incorporate the inelasticity reconstructions in the event binnings of previous IceCube oscillation analyses \cite{FLERCNN, QUESO}.  Higher inelasticity bins should be more neutrino pure due to the suppression of high inelasticity antineutrino events, thus helping to resolve the matter effects.  Correlation in the track classification scores (PID) and the inelasticity reconstructions must be considered.  For IC86 the events are separated into two inelasticity bins and each inelasticity bin is assigned its own PID binning.  For IC93 we apply the inelasticity binning only to the PID bins with a substantial amount of track events.  The binnings are shown in table~\ref{binning}.

\begin{equation}
\label{asimov_sensitivity}
\eta_\sigma = \frac{\Delta \chi_{NO-IO}(\mathrm{true\:ordering})-\Delta \chi_{NO-IO}(\mathrm{wrong\:ordering})}{2\sqrt{\Delta \chi_{NO-IO}(\mathrm{wrong\:ordering})}}
\end{equation}

Equation~\ref{asimov_sensitivity} is used to calculate the NMO sensitivities \cite{FLERCNN_NMO}.  We use a modified chi squared metric to account for limited MC statistics.  We calculate the sensitivities for a range of true values of $\theta_{23}$.  For comparison we also compute the sensitivities without using the inelasticity reconstructions.  The PISA software package is utilized for these calculations \cite{PISA}.  The IC86 sensitivities are computed assuming 9.28 years of livetime.  We utilize the same systematics and priors as used in a previous NMO study \cite{FLERCNN_NMO}.  The systematics account for uncertainties in the atmospheric neutrino and muon fluxes, cross sections, optical properties of the Antarctic ice, the detector, and the neutrino oscillation parameters.  For IC93 we assume a livetime of 3 years and choose to fix the systematic parameters to their nominal values and only minimize over $\theta_{23}$ and $\Delta m_{31}^2$.

\begin{figure}[htbp]
\centering
\includegraphics[width=.45\textwidth]{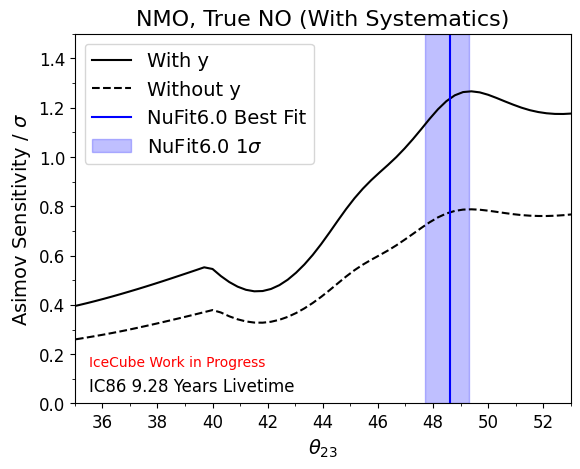}
\includegraphics[width=.45\textwidth]{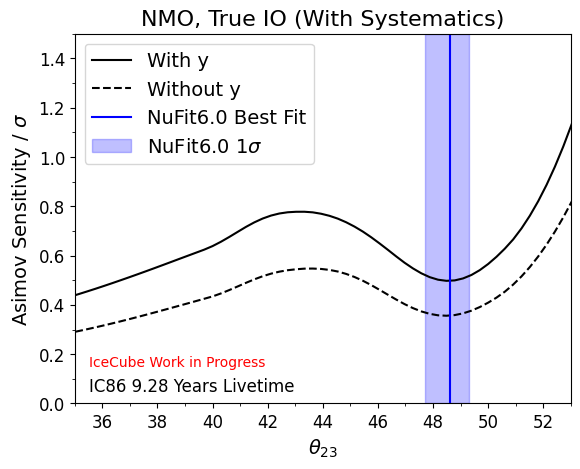}
\caption{\label{fig:ic86_sensitivities} Sensitivity to the NMO assuming the normal ordering (left) and inverted ordering (right) is true for IC86.}
\end{figure}

\begin{figure}[htbp]
\centering
\includegraphics[width=.45\textwidth]{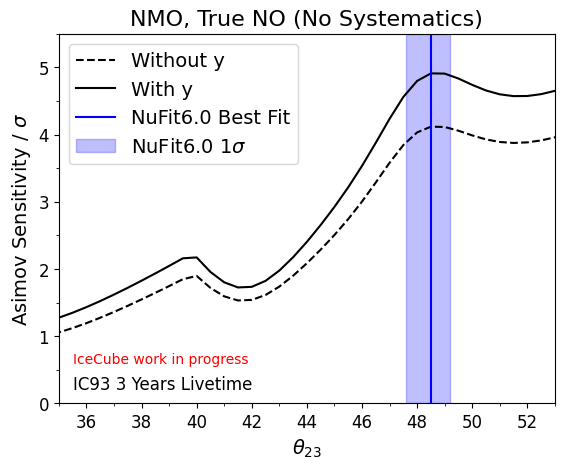}
\includegraphics[width=.45\textwidth]{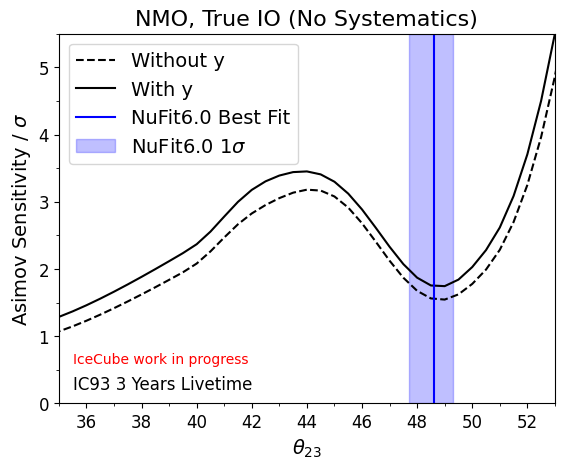}
\caption{\label{fig:ic93_sensitivities} Sensitivity to the NMO assuming the normal ordering (left) and inverted ordering (right) is true for IC93.}
\end{figure}

Figures~\ref{fig:ic86_sensitivities} and~\ref{fig:ic93_sensitivities} show the sensitivities obtained.  These show that incorporating inelasticity leads to an improvement in the NMO sensitivities for both IC86 and IC93, for both the normal and inverted orderings.  The largest improvement is seen for the normal ordering in the second octant of $\theta_{23}$, which coincides with the best fit value of $\theta_{23}$ from NuFIT 6.0 \cite{nufit}.

\section{Conclusions}
\label{sec:conc}

New inelasticity reconstruction algorithms have been developed for IceCube DeepCore and the IceCube Upgrade. Both have limited performance below 30 GeV, with significant improvement above 30 GeV.  Nevertheless, we find that incorporating inelasticity as a fourth observable increases the sensitivity to the neutrino mass ordering, for both mass orderings and detector configurations.  In the future we intend to improve the reconstruction algorithms and recalculate the IceCube Upgrade sensitivity including systematics.


\bibliographystyle{JHEP}
\bibliography{refs}

\end{document}